\documentclass[10pt, twocolumn, a4paper, fleqn]{article}

\usepackage{xcolor, fullpage, amsmath, amsfonts, amsgen, amssymb, amsbsy, graphicx, appendix, physics, hyperref, bm, bbm, palatino, flushend}
\usepackage[numbers,sort&compress]{natbib}
\usepackage{titlesec}
\titleformat{\section}{\fontsize{12}{12}\bfseries}{\thesection}{1em}{}
\titleformat{\subsection}{\fontsize{11}{11}\bfseries}{\thesubsection}{1em}{}
\titleformat{\subsubsection}{\fontsize{10}{10}\bfseries}{\thesubsubsection}{1em}{}

\hypersetup{colorlinks = true, linkbordercolor = white, linkcolor = blue, citecolor = blue}
\newcommand{\equa}[1]{Eq.~(\ref{#1})} \newcommand{\equas}[1]{Eqs.~(\ref{#1})}

\newcommand{\equasa}[2]{Eqs.~(\ref{#1}) and (\ref{#2})}
 
\newcommand{\Reyn}{{\rm Re}} 
\newcommand{\Ge}{{\rm Ge}}
 \newcommand{\Ra}{{\rm Ra}}  
\newcommand{\sdv}[1]{\frac{{\cal D}#1}{{\cal D} t}}
\newcommand{\str}{{\bm \tau}}
 
\newcommand{\eqn}[2]{\begin{gather}
#1
\label{#2}
\end{gather}
}

\title{\fontsize{14}{14}\bf The Boussinesq approximation for buoyant flows}

\author{{\fontsize{11}{11}\bf Antonio Barletta}\\[4pt]
{\small Department of Industrial Engineering. Alma Mater Studiorum Universit\`a di Bologna.}\\[-2pt] 
{\small Viale Risorgimento 2. 40136 Bologna. Italy}\\[-2pt]
{\small \tt antonio.barletta@unibo.it}}

\date{}

\begin{document}

\twocolumn[
\begin{@twocolumnfalse}
\maketitle
\begin{abstract}
\noindent 
The aim of this communication is to present a simplified, yet rigorous, deduction of the Boussinesq approximated governing equations for buoyant flows. In order to carry out the core deduction procedure, a simplified version of the manifold asymptotic analyses available in the literature is discussed. The method adopted in this study is focussed on the local balance equations valid for a general, not necessarily Newtonian, fluid. The analysis is carried out by demonstrating the leading order terms in the governing equations for the asymptotic limit which characterises the approximation. The role played by the effect of viscous dissipation is also taken into account.\vspace{1mm}
\end{abstract}
{\small \textbf{Keywords} --- Fluid flow; Convection; Buoyancy force; Boussinesq approximation; Pressure; Viscous dissipation.} 
\vspace{14mm}
  \end{@twocolumnfalse}
]

\section{Introduction}
There are several instances in fluid dynamics where an asymptotic analysis of the governing equations may lead to a simplified framework which offers an easier scheme for analytical or numerical solutions. Generally speaking, such asymptotic analyses lead to an approximation of the governing equations for the fluid flow. One of the most well-known asymptotic analyses is that leading to the boundary layer approximation widely employed for the study of external fluid flows and for the evaluation of the \mbox{fluid-solid} viscous interaction at the interface \cite{schlichting2003boundary}. Another example is the Boussinesq approximation for buoyancy-induced flows \cite{oberbeck1879warmeleitung, boussinesq1902mise, tritton1977physical, landau2013fluid, straughan2013energy, kundu2016fluid}. 

The Boussinesq approximation is the basis for the largest part of theoretical works on thermal convection published so far. The reliability of this approximation is mainly due to the extremely good agreement with experimental data. On the other hand, its foundation as an asymptotic theory has been laid out by several authors in quite complicated ways where the Boussinesq approximation turns out to be the result of a limit with multiple parameters tending to zero. 

This communication has not the aim of providing a comprehensive review of the really huge literature available on the topic. In fact, there are several excellent and comprehensive surveys in the literature \cite{spiegel1960boussinesq, mihaljan1962rigorous, gray1976validity, hillsroberts1991, rajagopal1996oberbeck, ZEYTOUNIAN2003575, rajagopal2009oberbeck, rajagopal2015approximation, grandi2021oberbeck, mayeli2021buoyancy}. The objective to be pursued in this paper is sketching a simplified, but rigorous, logical path that leads one from the general form of the local mass, momentum and energy balance equations to their approximate form. Strictly speaking, this aim is not completely original. In fact, the idea of the Boussinesq framework as the result of a limiting condition is already present in the literature. An example is the paper by \citet{gray1976validity}, where the limit is carried out starting from a Newtonian fluid with density, specific heat, viscosity, coefficient of thermal expansion and thermal conductivity dependent both on the temperature and on the pressure. The Boussinesq approximation is retrieved by defining eleven dimensionless \mbox{$\epsilon$-factors} which are considered as small. The complication in this approach, common also to other studies published before and after the paper by \citet{gray1976validity}, is in the lack of a precise and explicit definition of the physical assumptions at the bulk of the approximation scheme. Furthermore, the complicated mathematical framework sometimes tends to shade the physics underlying the analysis.

In this communication, the main purpose is capturing the chain of evidence supporting the Boussinesq approximation, by keeping the mathematical procedure as simple as possible and by declaring from the beginning the physical assumptions strictly needed to achieve the approximate governing equations. Despite the relatively simple structure of the presentation, no specific hypothesis is made for the Newtonian or non-Newtonian rheology of the fluid, or regarding the variability of fluid properties such as the viscosity or the thermal conductivity wherever not necessary. The density, as well as the pressure, the velocity and the temperature, is treated as a dynamic variable subject to specific assumptions. A brief discussion about the possible role of the viscous dissipation term in the local energy balance is also provided. 

An important point to be emphasised is that the limiting procedure leading to the Boussinesq approximation of the local balance equations for the fluid flow is an ad-hoc procedure. More precisely, the limit is taken in a specific way that leads exactly to those terms recognised a-priori as qualifying the approximation. This is neither questionable nor unexpected as every approximation is focussed on showing up that some specific terms in the balance equations turn out to be dominant over other terms, while further terms share the same order of magnitude. This is particularly evident if one thinks, for instance, to the boundary layer approximation for external flows \cite{schlichting2003boundary} where the ad-hoc assumption of a scale for the coordinate perpendicular to the bounding wall much smaller than the scale of the coordinate parallel to the wall is the basis to lay out the approximation.

\section{Governing equations}
Non-isothermal fluid flows are governed by the local mass, momentum and energy balance equations. We limit our analysis to cases where the buoyancy force typical of convection is due to the temperature differences. 
Mass diffusion and convection induced by the concentration differences of a solute can be treated in a completely similar manner and will not be considered here for the sake of simplicity. 

The governing equations of fluid flow can be expressed as \citep{landau2013fluid}
\eqn{
\pdv{\rho}{t} + \div\qty(\rho\, \vb{v}) = 0,\nonumber\\
\rho\, \sdv{\vb{v}} = - \grad P + \rho\, \vb{g} + \div \str,\nonumber\\
\rho\, \sdv{u} = - \div \vb{q} + \vb{D} : \str - P\, \div \vb{v} , 
}{1}
with the first \equa{1} yielding the local mass balance, the second the momentum balance and the third the energy balance. 

Here, $\rho$ is the density, $\vb{v}$ is the velocity, $t$ is time, $P$ is the pressure, $\vb{g}$ is the constant gravitational acceleration, $\str$ is the viscous stress tensor, $u$ is the internal energy per unit mass, $\vb{q}$ is the heat flux density and $\vb{D}$ is the strain tensor whose $ij$ component is given by
\eqn{
D_{ij} = \frac{1}{2} \, \qty(\pdv{v_i}{x_j} + \pdv{v_j}{x_i}), 
}{2}
with $x_i$ the $i$th Cartesian coordinate. 

Furthermore, the substantial derivative, ${\cal D}/{\cal D}t$, and the double scalar product between tensors, $\vb{D} : \str$, are defined as
\eqn{
\sdv{} = \pdv{}{t} + \vb{v} \vdot \grad , \qquad \vb{D} : \str = D_{ij}\, \tau_{ij} ,
}{3}
where repeated indices $ij$ are implicitly summed over. The term $\vb{D} : \str$ expresses the physical effect of viscous dissipation.

As a consequence of the first \equa{1}, the third \equa{1} can be rewritten as
\eqn{
\rho\, \qty[ \sdv{u} + P\, \sdv{}\qty(\frac{1}{\rho}) ] = - \div \vb{q} + \vb{D} : \str .
}{4}
By employing the thermodynamic equation for stable equilibrium states,
\eqn{
\dd u = T\, \dd s - P\, \dd \qty(\frac{1}{\rho}),
}{5}
where $s$ is the entropy per unit mass and $T$ is the thermodynamic temperature, we can write
\eqn{
\sdv{u} = T\, \sdv{s} - P\, \sdv{} \qty(\frac{1}{\rho}).
}{6}
Here, $1/\rho$ is the volume per unit mass. Thus, \equa{4} reads
\eqn{
\rho\, T\, \sdv{s} = - \div \vb{q} + \vb{D} : \str .
}{7}

\section{The Boussinesq scheme}
The Boussinesq approximation can be formulated as an asymptotic theory based on a system of five physical assumptions:
\begin{enumerate}
\item The fluid density depends only on the local temperature, so that pressure changes yield negligible density changes;
\item The local difference between the temperature and its reference constant value, $T_0$, is very small;
\item The acceleration experienced by the fluid elements in the flow domain is very small;
\item The viscous stress is very small;
\item The local difference between the pressure gradient and the hydrostatic pressure gradient is very small. 
\end{enumerate}
We will introduce and comment each of these assumptions, step by step, pointing out their effects.

\subsection{Small density variations}
\label{denvar}
We assume that the temperature varies very weakly both in time and in space. Hence, we can write
\eqn{
T - T_0 = \epsilon\, T',
}{7b}
where $\epsilon$ is a small perturbation parameter. Here and in the following, the primed quantities are implicitly assumed as being $\order{1}$ when, eventually, the limit of a vanishing $\epsilon$ is taken.

The density changes are induced only by temperature changes, while pressure variations yield only negligible effects.  
An alternative way to formulate this assumption is: ``any convective velocity is much smaller than the speed of sound in the fluid'' \cite{RevModPhys49581}.
Thus, we can write
\eqn{
\rho = \rho_0 \, \qty( 1 - \epsilon\, \theta) .
}{8}
In \equa{8}, $\rho_0$ is a constant expressing the reference fluid density and $\theta$ is a dimensionless function of $T'$ modelling the slight density changes in the fluid. More precisely, $\rho_0$ is the fluid density evaluated at the reference temperature $T_0$. The simplest and most common situation is when $\theta(T')$ is given by a linear function,
\eqn{
\theta(T) = \beta \, T' ,
}{9}
where $\beta$ is the thermal expansion coefficient. When \equa{9} is employed, $\epsilon$ has a direct physical meaning expressed as
\eqn{
\epsilon = \beta\, \Delta T_r,
}{10}
where $\Delta T_r$ is a constant reference temperature difference depending on the temperature boundary conditions imposed for the flow. 
A nonlinear expression of $\theta(T')$ may be needed to get a proper model of a density maximum, as it happens for water in a temperature range around $4\,^{\circ} {\rm C}$.

The role of $\epsilon$ is a qualifying point for the Boussinesq approximation, which can be formulated as an asymptotic regime achieved in the limit $\epsilon \to 0$.

\subsection{Local mass balance}
We rewrite the first \equa{1} by employing \equa{8},
\eqn{
- \rho_0\, \epsilon\, \pdv{\theta}{t} + \rho_0\, \div \vb{v} - \rho_0\, \epsilon \, \div\qty(\theta\, \vb{v})  = 0.
}{11}
By letting $\epsilon \to 0$, we obtain the usual condition of a solenoidal velocity field, namely
\eqn{
\div \vb{v} = 0.
}{12}
%

%
%
%

\subsection{Small acceleration and viscosity}\label{accvis}
Another qualifying point of the Boussinesq approximation is the small acceleration undergone in a buoyant flow, which means that the acceleration field
\eqn{
\vb{a} = \sdv{\vb{v}},
}{13}
can be scaled through the parameter $\epsilon$. Therefore, we can write
\eqn{
\vb{a} = \epsilon\, \vb{a}' .
}{14}
In fact, \citet{RevModPhys49581} explicitly state: ``any accelerations in the fluid are much less than $g$'', where $g$ is the modulus of $\vb{g}$. 

Due to the definition of substantial derivative, \equa{3}, the assumption of small acceleration can be mapped into one of large time and small velocity, through the scaling
\eqn{
\vb{v} = \epsilon^{1/2}\, \vb{v}' , \qquad t = \epsilon^{-1/2}\, t'.
}{15}
By using \equa{15}, \equas{3}, (\ref{13}) and (\ref{14}) in fact yield
\eqn{
\vb{a}' = \pdv{\vb{v}'}{t'} + \qty(\vb{v}' \vdot \grad) \vb{v}' .
}{16}
On account of \equa{15}, \equa{12} can be easily rewritten in terms of $\vb{v}'$,
\eqn{
\div \vb{v}' = 0.
}{12b}
The hypothesis of small viscous stress can be formulated as
\eqn{
\str = \epsilon \, \str' .
}{17}
The idea of small viscous stress as a characteristic feature of buoyant flows has a direct foundation in the argument usually set up to justify the onset of convection cells in a fluid layer heated from below \cite{RevModPhys49581}. In fact, the onset of convection emerges as a phenomenon where the viscous stress competes with the buoyancy force by damping out the initiation of the flow. Steady convection cells may arise only when the buoyancy force turns out to be dominant over the viscous stress \cite{RevModPhys49581}. 

We have not specified whether the fluid is Newtonian or not. Should the fluid be Newtonian, one can turn the hypothesis of a small viscous stress into one of small viscosity coefficient. Indeed, for a Newtonian fluid, 
we have
\eqn{
\tau_{ij} = 2\, \mu \, D_{ij}  ,
}{18}
where \equa{12} has been used and $\mu$ is the dynamic viscosity. 
Then, \equasa{15}{17} require that one must consistently assume
\eqn{
\mu = \epsilon^{1/2}\, \mu',
}{19}
so that one may write
\eqn{
\tau'_{ij} = 2\, \mu' \, D'_{ij} ,
}{20}
where $\vb{D}'$ is the tensor whose $ij$ component is given by
\eqn{
D'_{ij} = \frac{1}{2} \, \qty(\pdv{v'_i}{x_j} + \pdv{v'_j}{x_i}).
}{32}

\subsection{Hydrostatic pressure gradient}\label{hydpre}
The local momentum balance equation expressed by the second \equa{1} contains the gravitational body force $\rho\, \vb{g}$ which, on account of \equa{8}, can be written as
\eqn{
\rho\, \vb{g} = \rho_0\, \vb{g} - \rho_0\, \epsilon\, \theta\, \vb{g} .
}{21}
Hence, the difference between the pressure gradient, $\grad P$, and the gravitational body force, $\rho\, \vb{g}$, is given by
\eqn{
\grad{P} - \rho\, \vb{g} = \qty[ \grad P - \grad\qty(\rho_0\, \vb{g} \vdot \vb{x})] 
\nonumber\\ 
+ \rho_0\, \epsilon\, \theta\, \vb{g} ,
}{22}
where $\vb{x} = \qty(x_1, x_2, x_3)$ is the position vector. In \equa{22}, the term $\grad\qty(\rho_0\, \vb{g} \vdot \vb{x})$ is the hydrostatic pressure gradient. 

The last hypothesis that defines the basis of the Boussinesq asymptotic regime is relative to the difference between the pressure gradient and the hydrostatic pressure gradient. Such a difference is to be considered as extremely small everywhere. In mathematical terms, this hypothesis is formulated as
\eqn{
\grad P - \grad\qty(\rho_0\, \vb{g} \vdot \vb{x}) = \grad p, \qquad 
\nonumber\\ 
\grad p = \epsilon\, \grad{p'}.
}{23}
In \equa{23}, $\grad p$ is nothing but a shorthand notation for the difference between the pressure gradient and the hydrostatic pressure gradient. 
Equation~(\ref{23}) has very important physical consequences, because the hydrostatic pressure gradient is large or, at the very least, not negligible. Then, also the pressure gradient, $\grad P$, is a large or non-negligible vector quantity. This means that the pressure is markedly non-uniform in the vertical direction, despite the negligibly small compressibility effects on the density distribution. 

\subsection{Local momentum balance}
We now employ the assumptions laid out in Sections~\ref{denvar}, \ref{accvis} and \ref{hydpre} to rewrite the second \equa{1},
\eqn{
\rho_0\,\epsilon\, \qty[\pdv{\vb{v}'}{t'} + \qty(\vb{v}' \vdot \grad) \vb{v}'] 
\nonumber\\ 
- \rho_0\,\epsilon^2 \theta \qty[ \pdv{\vb{v}'}{t'} + \qty(\vb{v}' \vdot \grad) \vb{v}'] 
\nonumber\\
= - \epsilon\, \grad p' - \rho_0\, \epsilon\, \theta\, \vb{g} + \epsilon\, \div \str'.
}{24}
We divide \equa{24} by $\epsilon$, we simplify and then we let $\epsilon \to 0$. Thus, we obtain
\eqn{
\rho_0\, \qty[\pdv{\vb{v}'}{t'} + \qty(\vb{v}' \vdot \grad) \vb{v}'] 
\nonumber\\ 
= - \grad p' - \rho_0\, \theta\, \vb{g} + \div \str'.
}{25}
As pointed out in Section~\ref{denvar}, in most buoyant flows, \equa{9} holds. Then, \equa{25} can be rewritten as
\eqn{
\rho_0\, \qty[\pdv{\vb{v}'}{t'} + \qty(\vb{v}' \vdot \grad) \vb{v}'] 
\nonumber\\ 
= - \grad p' - \rho_0\, \beta \, T'\, \vb{g} + \div \str'.
}{26}
The term $- \rho_0\, \beta \, T'\, \vb{g}$ is the buoyancy force.

\subsection{Local energy balance}
The local energy balance equation given by \equa{7} can be rewritten by employing the results collected in Sections~\ref{denvar}, \ref{accvis} and \ref{hydpre}. The first consideration is relative to the term $T \, {\cal D} s/{\cal D} t$ on the left hand side of \equa{7}. By utilising the thermodynamic definitions of specific heat at constant volume and of specific heat at constant pressure,
\eqn{
c_v = T\, \qty(\pdv{s}{T})_{\!\!\rho}, \qquad c_P = T\, \qty(\pdv{s}{T})_{\!\!P},
}{27}
one can write
\eqn{
T\, \dd s = c_v\, \dd T,
}{28}
for a constant density process. On the other hand, one has
\eqn{
T\, \dd s = c_p\, \dd T,
}{29}
for a constant pressure process. The question whether the Boussinesq approximation is described effectively by an isobaric process of the elementary fluid element or by an isochoric process can be easily responded by recognising that:
\begin{itemize}
\item The density changes are $\order{\epsilon}$ as revealed by \equa{8};
\item The pressure changes in the vertical direction are as large as the hydrostatic pressure changes as revealed by \equa{23}.
\end{itemize}
Convection cells display a nonvanishing vertical component of velocity which means a sensible pressure variation in the process undergone by the fluid element. Under these conditions, the use of \equa{28} is the most satisfactory basis for the Boussinesq approximation, which becomes even an exact statement in the case of perfect gases. This can be easily reckoned by inspecting the third \equa{1} after the substitution of \equa{12}, and by recalling that in the case of a perfect gas we can write $\dd u = c_v\, \dd T$, whatever is the type of process undergone by the fluid element. If \equa{28} looks like the most reliable basis for the Boussinesq approximation, it cannot be retained as exact in general. We will thus use the symbol $c$ for the specific heat to be employed in the Boussinesq approximation having in mind that its value is to be established experimentally, though it can be well estimated by equating $c$ with $c_v$. Then, we rewrite \equa{7} as
\eqn{
\rho\, c\, \sdv{T} = - \div \vb{q} + \vb{D} : \str .
}{30}
We now use \equasa{7b}{15}, together with \equa{17}, to conclude that
\eqn{
\rho\, c\, \sdv{T} = \epsilon^{3/2}\, \rho_0\, c\, \qty( \pdv{T'}{t'} + \vb{v}' \vdot \grad T' )
\nonumber\\ 
- \epsilon^{5/2}\, \rho_0\, c\, \theta\, \qty( \pdv{T'}{t'} + \vb{v}' \vdot \grad T' ),
\nonumber\\
\vb{D} : \str = \epsilon^{3/2}\, \vb{D}' : \str' ,
}{31}
%
The heat flux density is expressed through Fourier's law, so that
\eqn{
\vb{q} = - \kappa\, \grad T ,
}{33}
where $\kappa$ is the thermal conductivity of the fluid. 

\subsubsection{Non-negligible viscous dissipation}\label{visdis}
From \equa{33}, the diffusion term, $- \div \vb{q}$, in \equa{30} is $\order{\epsilon^{3/2}}$, as the other terms expressed by \equa{31}, if
\eqn{
\kappa = \epsilon^{1/2}\, \kappa' .
}{34}
Thus, we can write
\eqn{
- \div \vb{q} = - \epsilon^{3/2} \, \div \vb{q}' ,
}{35}
where
\eqn{
\vb{q}' = - \kappa'\, \grad T' .
}{36}
By substituting \equasa{31}{35} into \equa{30}, we obtain
\eqn{
\epsilon^{3/2}\, \rho_0\, c\, \qty( \pdv{T'}{t'} + \vb{v}' \vdot \grad T' ) 
\nonumber\\ 
- \epsilon^{5/2}\, \rho_0\, c\, \theta\, \qty( \pdv{T'}{t'} + \vb{v}' \vdot \grad T' ) 
\nonumber\\
= - \epsilon^{3/2} \, \div \vb{q}' + \epsilon^{3/2}\, \vb{D}' : \str' .
}{37}
We divide \equa{37} by $\epsilon^{3/2}$ and then we take the limit $\epsilon \to 0$. The resulting local energy balance equation is
\eqn{
\rho_0\, c\, \qty( \pdv{T'}{t'} + \vb{v}' \vdot \grad T' ) 
\nonumber\\
= - \div \vb{q}' +  \vb{D}' : \str' .
}{38}
Here, there is an implicit assumption that $c$ is $\order{1}$. This assumption, together with that expressed by \equa{34}, may have a possible alternative.

\subsubsection{Negligible viscous dissipation}
\label{nonvisdis}
On account of \equa{33}, the diffusion term, $- \div \vb{q}$, in \equa{30} is $\order{\epsilon}$ with
\eqn{
- \div \vb{q} = - \epsilon \, \div \vb{q}' ,
}{39}
if we introduce the alternative statement
\eqn{
\vb{q}' = - \kappa\, \grad T' ,
}{40}
where $\kappa$ is considered $\order{1}$. We assume
\eqn{
c = \epsilon^{-1/2} \, c' ,
}{41}
instead, which depicts a situation where the fluid has a large specific heat.
We now substitute \equas{31}, (\ref{39}) and (\ref{41}) into \equa{30}, so that we obtain
\eqn{
\epsilon\, \rho_0\, c'\, \qty( \pdv{T'}{t'} + \vb{v}' \vdot \grad T' ) 
\nonumber\\ 
- \epsilon^{2}\, \rho_0\, c'\, \theta\, \qty( \pdv{T'}{t'} + \vb{v}' \vdot \grad T' ) 
\nonumber\\
= - \epsilon \, \div \vb{q}' + \epsilon^{3/2}\, \vb{D}' : \str' .
}{42}
We divide \equa{42} by $\epsilon$ and then we take the limit $\epsilon \to 0$. The resulting local energy balance equation is
\eqn{
\rho_0\, c'\, \qty( \pdv{T'}{t'} + \vb{v}' \vdot \grad T' ) 
= - \div \vb{q}' .
}{43}
Unlike \equa{38}, \equa{43} does not include any contribution due to viscous dissipation.

\section{Analysis of the approximation}
We have drawn two different paths in Sections~\ref{visdis} and \ref{nonvisdis} leading to the approximate local energy balance expressed by either \equa{38} or \equa{43}. If viscous dissipation turns out to be non-negligible in \equa{38}, this effect is neglected in \equa{43}. The choice between the two alternatives relies on the identification of the appropriate scenario. We can devise a fluid with a very small thermal conductivity where viscous dissipation may be important and \equa{38} is the appropriate Boussinesq approximation of the energy balance. Alternatively, we can devise a fluid with a very large specific heat where the effect of viscous dissipation is negligibly small and \equa{43} is the appropriate energy balance. The physics behind this mathematical result can be focussed on by considering the Gebhart number, a nondimensional parameter often employed for the analysis of viscous dissipation in natural or mixed convection flows,
\eqn{
\Ge = \frac{g\, \beta\, L}{c},
}{44}
where $L$ is a typical length characterising the flow domain. If we follow the path drawn in Section~\ref{visdis}, $c$ is $\order{1}$ so that also $\Ge$ is $\order{1}$. If we follow the path drawn in Section~\ref{nonvisdis}, $c$ is $\order{\epsilon^{-1/2}}$ so that also $\Ge$ is $\order{\epsilon^{1/2}}$ and, hence, $\Ge \ll 1$. 

These findings are consistent with the usual idea that viscous dissipation is negligible in a buoyant flow when the Gebhart number is extremely small. An interesting aspect of the alternative views described in Sections~\ref{visdis} and \ref{nonvisdis} is that both views imply a thermal diffusivity of the fluid,
\eqn{
\alpha = \frac{\kappa}{\rho_0\, c},
}{45}
which turns out to be $\order{\epsilon^{1/2}}$,
\eqn{
\alpha = \epsilon^{1/2} \, \alpha' .
}{46}
This is a very important point, as the thermal diffusivity is involved in the definition of Prandtl number,
\eqn{
\Pr = \frac{\mu}{\rho_0\, \alpha} ,
}{47}
and \equasa{19}{46} allow one to write
\eqn{
\Pr = \frac{\mu'}{\rho_0\, \alpha'} .
}{48}
Equation~(\ref{48}) means that a finite Prandtl number corresponds to either choices proposed in Sections~\ref{visdis} and \ref{nonvisdis}. 

Another sensible dimensionless parameter for convection is the Rayleigh number, which is defined as
\eqn{
\Ra = \frac{\rho_0\, g\, \beta\, \Delta T_r\, L^3}{\mu\, \alpha} .
}{49}
By employing \equas{10}, (\ref{19}) and (\ref{46}), \equa{49} can be rewritten as
\eqn{
\Ra = \frac{\rho_0\, g\, \epsilon\, L^3}{\epsilon\, \mu'\, \alpha'} = \frac{\rho_0\, g\, L^3}{\mu'\, \alpha'} ,
}{50}
which means that the Rayleigh number is $\order{1}$. A similar result is drawn for the Reynolds number, which is an important parameter in a regime of mixed convection. Its definition is
\eqn{
\Reyn = \frac{\rho_0\, V_r\, L}{\mu} ,
}{51}
where $V_r$ is a constant expressing the reference velocity of the forced flow. Such a constant undergoes the same behaviour of every local value of velocity, expressed by \equa{15}, so that we have
\eqn{
V_r = \epsilon^{1/2}\, V'_r .
}{52}
By using \equasa{19}{52}, we can rewrite \equa{51} as
\eqn{
\Reyn = \frac{\rho_0\, \epsilon^{1/2}\, V'_r\, L}{\epsilon^{1/2}\, \mu'} = \frac{\rho_0\, V'_r\, L}{\mu'},
}{53}
meaning that the Reynolds number is $\order{1}$.

The approximate local balance equations (\ref{12b}), (\ref{26}) and (\ref{38}), or alternatively (\ref{43}), are expressed through primed quantities which means that such equations are upscaled with the respect to their natural units. Such a formulation has been functional to achieving the limit $\epsilon \to 0$ by selecting the dominant terms in each balance equation. Now, one can restore the natural scales of temperature, velocity and pressure by multiplying \equa{12b} by $\epsilon^{1/2}$, \equa{26} by $\epsilon$, \equa{38} by $\epsilon^{3/2}$, or \equa{43} by $\epsilon$. This operation allows one to retrieve the usual physical scales for the fluid dynamics variables, so that the Boussinesq approximation yields
\eqn{
\div{\vb{v}}=0,
\nonumber\\
\rho_0\, \qty[\pdv{\vb{v}}{t} + \qty(\vb{v} \vdot \grad) \vb{v}] 
\nonumber\\ 
= - \grad p - \rho_0\, \beta \, \qty(T - T_0) \vb{g} + \div \str,
\nonumber\\
\rho_0\, c\, \qty( \pdv{T}{t} + \vb{v} \vdot \grad T ) 
\nonumber\\
= - \div \vb{q} +  \vb{D} : \str , \qif \Ge \sim \order{1} ,
}{54}
or, as an alternative to the third \equa{54},
\eqn{
\rho_0\, c\, \qty( \pdv{T}{t} + \vb{v} \vdot \grad T ) 
\nonumber\\
= - \div \vb{q} , \qif \Ge \ll 1 .
}{55}

\section{Conclusions}
The Boussinesq approximation, widely employed for the theoretical modelling of buoyant flows either in natural convection or in mixed convection has been discussed. The basic physical assumptions leading to this approximate framework have been explicitly declared from the beginning and the limiting procedure has been described step-by-step in a form simplified with respect to previous papers. Unlike in other theoretical studies on the topic, no explicit assumption is made about the fluid rheology or the variability of fluid properties such as the specific heat, the coefficient of thermal expansion or the thermal conductivity. The effect of viscous dissipation can be considered as important when the fluid thermal conductivity is a small parameter. This feature, within the Boussinesq approximation, captures the physics of the viscous dissipation role in the description of buoyant flows. When the fluid is a poor heat conductor, the internal frictional heating has a chance to influence the temperature distribution in a significant manner. On the other hand, this is an unlikely case if the fluid has a large thermal conductivity.

\section*{Acknowledgements}
The author acknowledges the financial support from grant PRIN~2017F7KZWS provided by the
Italian Ministry of Education, University and Research.

\end{document}